\documentclass{article}
\usepackage{spconf,amsmath,graphicx, amssymb}
\usepackage{amsfonts,bm}
\usepackage{mathtools}
\usepackage{soul}
\usepackage{color}

\newcommand{\Amat}{\mathbf{A}_{\bm{(\Phi)}}}
\newcommand{\avec}{\mathbf{a}_m}
\newcommand{\Fmat}{\mathbf{F}}
\newcommand{\xvec}{\mathbf{x}}
\newcommand{\yvec}{\mathbf{y}}
\newcommand{\svec}{\mathbf{s}}
\newcommand{\prox}{\mathcal{P}_{(\zeta)}}
\newcommand{\stepsize}{\alpha_{(\psi)}}
\newcommand{\phimn}{\phi_{m,n}}
\newcommand{\phivec}{\boldsymbol{\phi}_{m}}
\newcommand{\pimn}{\pi_{m,n}}
\newcommand{\evec}{\mathbf{e}_{m}}
\newcommand{\emn}{e_{m,n}}

\newcommand{\Dent}{D_{\xi}}
\newcommand{\ffunc}{f_{\theta}}

\newcommand{\RealN}{\mathbb{R}^{N}}

\newcommand{\CompN}{\mathbb{C}^{N}}
\newcommand{\CompM}{\mathbb{C}^{M}}
\DeclarePairedDelimiter\norm{\lVert}{\rVert}
\title{Learning Sampling and Model-Based Signal Recovery  \\ for Compressed Sensing MRI}
\name{Iris A.M. Huijben$^1$, Bastiaan S. Veeling$^2$, and Ruud J.G. van Sloun$^1$
\thanks{This research was supported in part by Philips Research. It is also part of a research program Rubicon ENW 2018-3 with project number 019.183.EN.014, which is financed by the Dutch Research Council (NWO). We thank Wouter Kool and the anonymous reviewers for their valuable feedback.}}
\address{$^1$Dept. of Electrical Engineering, Eindhoven University of Technology, The Netherlands \\
$^2$Dept. of Computer Science, University of
Amsterdam, The Netherlands}

\begin{document}

%
\maketitle
\begin{abstract}
Compressed sensing (CS) MRI relies on adequate undersampling of the k-space to accelerate the acquisition without compromising image quality. Consequently, the design of optimal sampling patterns for these k-space coefficients has received significant attention, with many CS MRI methods exploiting variable-density probability distributions. Realizing that an optimal sampling pattern may depend on the downstream task (e.g. image reconstruction, segmentation, or classification), we here propose joint learning of both task-adaptive k-space sampling and a subsequent model-based proximal-gradient recovery network. The former is enabled through a probabilistic generative model that leverages the Gumbel-softmax relaxation to sample across trainable beliefs while maintaining differentiability. The proposed combination of a highly flexible sampling model and a model-based (sampling-adaptive) image reconstruction network facilitates exploration and efficient training, yielding improved MR image quality compared to other sampling baselines. 

\end{abstract}
\section{Introduction}
\label{sec:intro}

Magnetic resonance imaging (MRI) is an invaluable non-invasive medical imaging modality that enables reliable diagnostic imaging for a wide range of clinical applications with unmatched soft-tissue contrast and high resolution. It is however also associated with long acquisition times, thereby diminishing patient comfort, compromising high-quality dynamic imaging (e.g. cardiac), and yielding high procedural costs with limited throughput. To overcome this, accelerated MRI approaches have extensively been studied, initially resulting in parallel imaging methods that are commonplace in today's MRI devices, such as SENSE \cite{pruessmann1999sense}. 

More recently, methods that exploit both redundancy in the spatial frequency components as well as structural image priors (e.g. sparsity in some domain) have been proposed. These compressed sensing (CS) approaches \cite{donoho2006compressed,eldar2012compressed,candes2006robust} speed-up data acquisition by undersampling the full k-space through hand-tailored sampling schemes, while retaining image quality by exploiting the aforementioned sparsity during image reconstruction. Popular sampling schemes follow from variable-density probability density functions (PDF) \cite{lustig2007sparse}, which are mostly based on the empirical observation that lower frequencies should be sampled more densely than high frequencies to enable adequate image recovery. The authors of \cite{adcock2017breaking} bridge the gap between these empirically found variable-density sampling designs and the mathematical justification. In practice, optimal k-space sampling may however depend on the anatomy and the imaging task at hand (e.g whole body, dynamic cardiac, or segmentation), and these approaches therefore fail to exploit the full data distribution available. 


With the aim of exploiting such information, several data-driven approaches have been introduced for optimization of k-space sampling. Among these, greedy algorithms were proposed \cite{jin2019self,sanchez2019scalable} to handle the factorial scaling of this problem. Other approaches leverage gradient-based learning of sampling schemes through compact (gradient-distributing) interpolation kernels \cite{weiss2019pilot}. This was however found to inhibit effective exploration and flexibility, with learned subsampling schemes not deviating far from their initialization. 

In pursuit of a more effective and flexible approach to joint learning of sampling and image recovery, we here propose to adopt a generative k-space sampling model \cite{huijben2019learning}, termed Deep Probabilistic Subsampling (DPS), along with a concurrent unfolded proximal gradient network \cite{parikh2014proximal} for recovery. DPS leverages the recently proposed Gumbel-softmax relaxation for differentiable sampling from categoricals and distributes trainable beliefs over relevant k-space coefficients. The unfolded recovery model exploits both the known measurement domain transform (Fourier) and the sampling itself, while learning an effective image proximal mapping. This allows for fast co-adaptation of the reconstruction network to the selected samples and promotes exploration of an optimal sampling pattern. 

The remainder of this paper is structured as follows. We start by giving a description of the system model in section \ref{sec:systemModel}, after which we elaborate on our subsampling and reconstruction methods in sections \ref{sec:Sampling} and \ref{sec:Reconstruction}, respectively. Section \ref{sec:Experiments} elucidates upon the conducted experiments: sparse signal recovery from partial Fourier measurements (\ref{sec:CaseA}), and MRI reconstruction from partial k-space measurements (\ref{sec:CaseB}). We conclude in section \ref{sec:conclusion}.

Throughout the paper, bold capital letters denote matrices. We index matrices by row-first notation, i.e. $\avec$ denotes the $m^{\text{th}}$ row-vector of matrix $\mathbf{A}$, of which $a_{m,n}$ is the scalar in the $n^{\text{th}}$ column.


\section{Methods}
\label{sec:Methods}

\subsection{System model}
\label{sec:systemModel}
We define a partial Fourier measurement $\yvec \in \CompM$, acquired by subsampling
the full vector of Fourier coefficients $\xvec \in \CompN$, using $\Amat~\in~\{0,1\}^{M \times N}$, 
parameterized by $\Phi$:
\begin{equation}
    \yvec = \Amat\xvec.
\end{equation}
For each row $\avec$ in $\Amat$ we require $\norm{\avec}_1 = 1$, making $\Amat$ a subset selector that selects $M$ out of $N$ ($M \ll N$) elements from $\xvec$. We aim to achieve a subsequent task $\svec$ (e.g. image reconstruction) from $\yvec$ through a nonlinear task function $\ffunc(\cdot)$ that is differentiable with respect to its trainable parameters $\theta$:
\begin{equation}
    \hat{\svec} = \ffunc(\yvec).
\end{equation}
%

\subsection{DPS: Deep probabilistic subsampling}
\label{sec:Sampling}
To enable training of both the parameters for the subsampling scheme, $\boldsymbol{\Phi}$, and the task network, $\theta$, we adopt a probabilistic generative sampling model that circumvents the non-differentiable nature of subset selection $\cite{huijben2019learning}$. Rather than attempting to train the sampling directly, we instead update beliefs across all possible Fourier (or k-space) coefficients.



To that end, we introduce $M$ independent categorical random variables, following distributions $\mathrm{Cat}_m(N, \boldsymbol{\pi}_m)$, with $m \in \{1,\ldots,M\}$, that all express beliefs across the full set of $N$ coefficients though the $N$ normalized class probabilities $\pi_{m,n}$ in $\boldsymbol{\pi}_m \in\RealN$. We define a logit $\phimn$ as the natural logarithm of the unnormalized class probability, such that:
\begin{equation}
\label{eqn:pimnDef}
\pimn=\frac{\mathrm{exp}~\phimn}{\sum_{i=1}^{N} \mathrm{exp}~\phi_{m,i}}.
\end{equation}

To draw a sample from a categorical, we exploit the Gumbel-max trick \cite{gumbel1954statistical}. This trick perturbs the unnormalized logits $\phivec$ with Gumbel noise $\evec$ in order to create randomized keys from which the highest key is selected. Subsequently a length-$N$ one-hot vector is created, i.e. a vector that contains only one non-zero element (with value 1) at the index of the drawn sample, through a function that we denote as $\mathrm{onehot}_N(\cdot)$. We iteratively draw samples without replacement across the $M$ categoricals, which we implement by masking previously sampled categories (here, Fourier coefficients) through $\mathbf{w}_{m-1} \in \{-\infty, 0\}$. Each row of $\Amat$ thus takes the following form:
%
\begin{equation}
    \label{eqn:onehotSample}
    \avec = \mathrm{onehot}_N\Big\{\underset{n \in \{1\ldots N\}}{\mathrm{argmax}}\big\{w_{m-1,n}+\phimn+\emn\big\}\Big\}.
\end{equation}    

To enable error backpropagation, we adopt the Gumbel-softmax trick, thereby relaxing the non-differentiable $\mathrm{argmax}(\cdot)$ function (hard sampling) by replacing it with the $\mathrm{softmax}_{\tau}(\cdot)$ function (termed soft sampling) \cite{jang2017categorical,maddison2016concrete}, having temperature parameter $\tau$. We thus define:
\begin{align}
    \label{eqn:gradAm}
    &\nabla_{\phivec}\ \avec \coloneqq 
    \nabla_{\phivec} \mathbb{E}_{\evec}\big[\mathrm{softmax}_{\tau}\{\mathbf{w}_{m-1}+\phivec+\evec\}\big] \nonumber =\\
     & \nabla_{\phivec} \mathbb{E}_{\evec}\Bigg[\frac{\mathrm{exp}\{(\mathbf{w}_{m-1}+\phivec +
    \evec)\mathrm{/}\tau\}}{\sum_{i=1}^{N}\mathrm{exp}\{(w_{m-1,i}+\phi_{m,i} + e_{m,i})\mathrm{/}\tau\}}\Bigg].
\end{align}
Across all experiments presented in this paper, the unnormalized logits $\boldsymbol{\Phi}$ are initialized by i.i.d zero-mean Gaussian noise ($\sigma^2=\frac{1}{4}$) samples. \\

\noindent\textbf{Top-\textit{M} sampling for large \textit{N}}\hspace{0.5cm} Defining $M$ categorical random variables with $N$ categories offers high expressiveness, but comes at the cost of having a large amount of trainable parameters and poses memory challenges. This particularly holds for large $N$, e.g. when sampling MRI k-space coefficients.
As such, we also consider sampling $M$ elements without replacement from only a single categorical distribution with trained logits, termed top-$M$ sampling. 

As shown by Kool \textit{et al.} \cite{kool2019stochastic}, top-$M$ hard sampling is equivalent to sampling $M$ times without replacement from the same distribution. As such, we can define
\begin{align}
    \label{eqn:MhotSample}
    \Amat
    &= \mathrm{Mhot}_N\big\{\underset{n \in \{1\ldots N\}}{\mathrm{topM}}(\phi_n+e_n)\big\},
\end{align}
where $\mathrm{topM}(\cdot)$ returns the indices of the unique $M$ highest values, and $\mathrm{Mhot}_N(\cdot)$ creates $M$ one-hot encoded rows from the returned indices by $\mathrm{topM}(\cdot)$, yielding the rows of $\Amat$.

The authors of \cite{xie2019reparameterizable} demonstrated that iterative sampling without replacement from a single Gumbel-softmax distribution is a valid top-$M$ relaxation. As such we can directly leverage \eqref{eqn:gradAm} for backpropagation. 

\subsection{Model-based deep networks for signal recovery}
\label{sec:Reconstruction}

In CS, the ill-posed problem of signal/image reconstruction from compressive measurements is typically solved using proximal gradient methods \cite{parikh2014proximal}. The iterative nature makes them tediously slow however, and the adopted regularizers are typically selected empirically. Recently, the use of deep neural networks for signal/image reconstruction has gained popularity, being a fast, fully data-driven alternative. Common network designs are closely related to the U-Net \cite{ronneberger2015u,bahadir2019learning}. 
While popular, such a structure is inherently not sampling-adaptive and does not exploit available knowledge on the acquisition. We therefore propose to leverage a deep unfolded proximal gradient method \cite{parikh2014proximal}, in which iterations of a known iterative algorithm are unfolded as a feedforward neural network. The update rule of a this proximal gradient scheme takes the following form: 
%
\begin{align}
    \label{eqn:ISTAupdate}
    \hat{\svec}^{(k+1)} &= \prox^{(k)}\left\{\hat{\svec}^{(k)}-\stepsize^{(k)}\Fmat^{H}\Amat^T\left(\Amat\Fmat\hat{\svec}^{(k)}-\Amat\xvec\right)\right\} \\ 
    \label{eqn:ISTAupdate2}
    &= \prox^{(k)}\left\{B^{(k)}\hat{\svec}^{(k)} +C^{(k)}[\Amat^T\Amat\xvec]\right\},
\end{align} 
where $\Fmat \in \mathbb{C}^{N \times N}$ is the Fourier transform matrix, $\Fmat^{H}$ is its Hermitian, $\alpha^{(k)}$ and $\prox(\cdot)$ are the stepsize and the proximal operator (parameterized by $\zeta$), respectively, $B^{(k)} = \mathbf{I}-\stepsize^{(k)}\Fmat^{H}\Amat^{T}\Amat\Fmat$, and $C^{(k)} = \stepsize\Fmat^{H}$. Each iteration in \eqref{eqn:ISTAupdate} is a model-based (MB) network layer, with trainable parameters $\zeta$. Depending on the application, we will either fully learn a suitable proximal mapping $\prox(\cdot)$ using a neural network (MRI reconstruction), or only learn a layer-wise shrinkage parameter for a known analytic proximal operator (sparse recovery). 

This MB-approach not only facilitates the incorporation of domain transforms and sampling patterns in the network architecture (allowing it to co-adapt when learning the sampling scheme), but also greatly reduces the amount of trainable parameters compared to conventional convolutional networks.

\section{Experiments}
\label{sec:Experiments}

\subsection{Partial Fourier sampling of sparse signals}
\label{sec:CaseA}

\textbf{Experiment setup}\hspace{0.5cm} 
We first demonstrate DPS for sparse signal reconstruction from partial Fourier measurements \cite{lustig2007sparse,lustig2008compressed,zbontar2018fastmri}. To that end, we synthetically generate random sparse signal vectors $\svec\in\mathbb{R}^{128}$, with 5 non-zero coefficients, which we subsequently Fourier-transform to yield $\mathbf{x}\in\mathbb{C}^{128}$.
We compare subsampling using a fixed random pattern (classic CS) to DPS, using either top-$M$ or top-1 sampling. 
\vspace{0.2cm}

\noindent\textbf{Reconstruction model}\hspace{0.5cm} 
The MB recovery network is based on an unfolded version of the iterative shrinkage and thresholding algorithm \cite{gregor2010learning}. It comprises 3 layers that follow the iterations in \eqref{eqn:ISTAupdate2}, with $B^{(k)}$ and $C^{(k)}$ here replaced by trainable fully-connected layers, and  $\prox(\cdot)^{(k)}$ being a smoothed sigmoid-based soft-thresholding operator \cite{atto2008smooth}, with learned thresholding parameters for each $k$. We compare this MB network to a standard fully-connected (FC) network with 5 hidden layers containing $256, 512, 256, 128$, and $128$ nodes, respectively, each followed by leaky ReLU activations.


\vspace{0.2cm}

\noindent\textbf{Training details}\hspace{0.5cm}
\label{sec:TrainingDetailsCaseA}
The model is trained by minimizing the mean-squared error (MSE) \label{eqn:Ls} between the model predictions and the ground truth signals $\svec$. 
For DPS, we promote training towards one-hot distributions $\mathrm{Cat}_m(N, \boldsymbol{\pi}_m)$ through an additional mild entropy penalty (multiplier $\mu = 1e-8$) on the trainable logits.
Note that this penalty is not applied in case of top-$M$ sampling. 
%
The resulting total loss is minimized using stochastic gradient descend across mini-batches of 16 randomly-generated Fourier-transformed data vectors. To this end we adopt the ADAM solver ($\beta_1 = 0.9, \beta_2 = 0.999$, and $\epsilon = 1e{-7}$) \cite{kingma2014adam}, with separate learning rates for the generative sampling model and the reconstruction network ($8{e-2}$ and $1{e-3}$, respectively).  We train until the validation loss converged, with the temperature parameter $\tau$ gradually lowered from $5.0$ to $0.5$. \vspace{0.2cm}

\begin{figure}
\centering
\includegraphics[width=\linewidth]{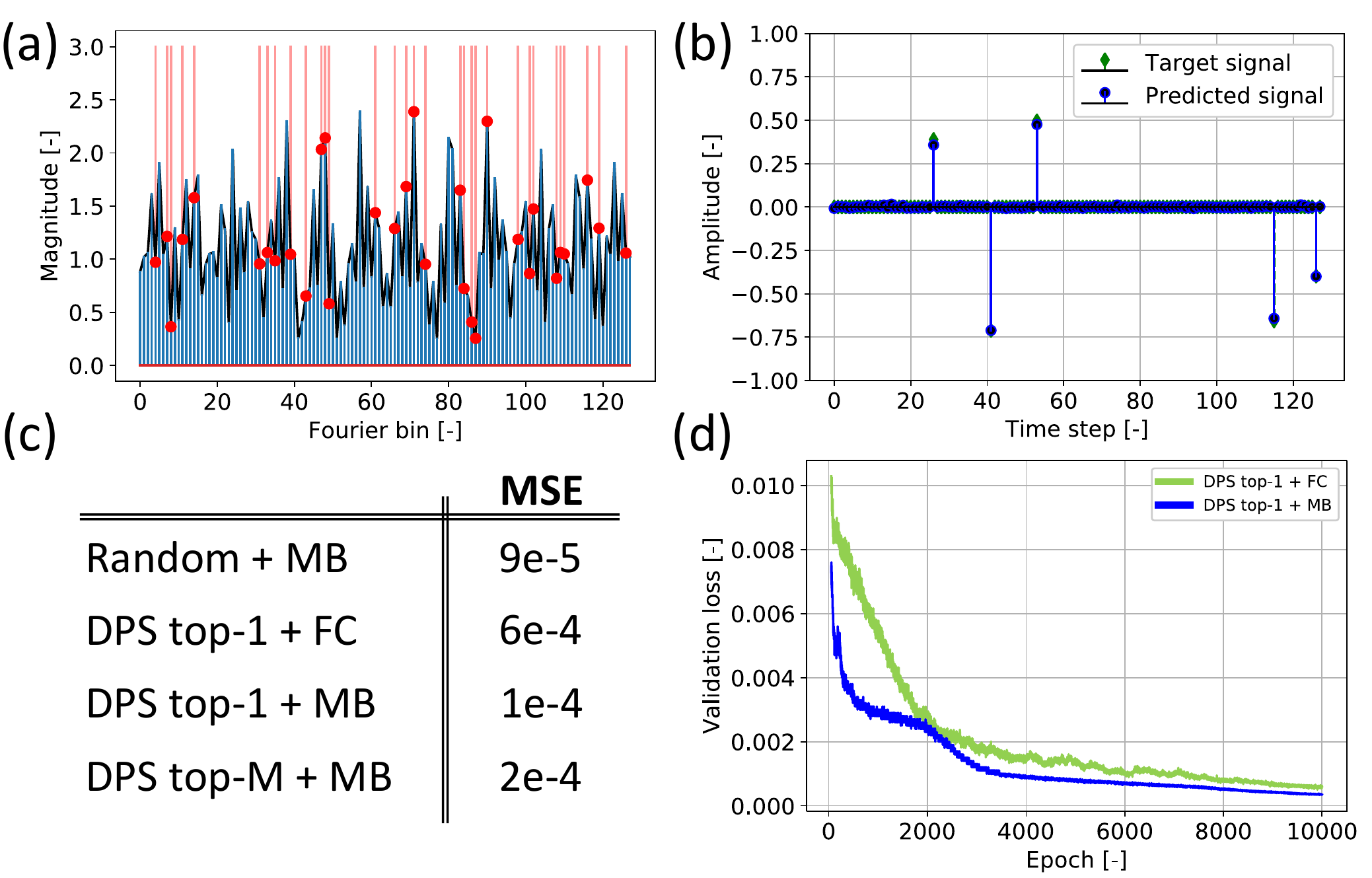}
\caption{\small (a) Partial Fourier measurements by Deep Probabilistic Subsampling (DPS) with $N/M=4$ (red) and all Fourier coefficients (blue). (b) Reconstructed sparse signal using the proposed model-based (MB) deep network. (c) MSE values for DPS vs random partial Fourier measurements, and MB vs fully-connected (FC) reconstruction networks. (d) Training graphs for DPS top-1 with MB (blue) and FC (green) networks for the first 10,000 epochs.}
\label{fig:FourierRecon}
\end{figure}

\noindent\textbf{Results}\hspace{0.5cm}
Figure \ref{fig:FourierRecon}b visualizes sparse signal reconstruction using the proposed MB deep network with learned subsampling by DPS (top-1 sampling), selecting $25\%$ of the Fourier coefficients (see fig. \ref{fig:FourierRecon}a, samples are displayed in red). The sparse signal components are well-recovered. Figure \ref{fig:FourierRecon}c provides the MSE values for all tested models, and the training graphs (first 10,000 epochs) for DPS top-1 with MB and fully-connected recovery networks are displayed in fig. \ref{fig:FourierRecon}d. We clearly see the positive effect of a MB recovery network compared to a FC network in terms of performance and convergence, which we attribute to the MB-network's ability to efficiently co-adapt with the changing sampling pattern. 

\subsection{MR images from partial k-space measurements} 
\label{sec:CaseB}

\begin{figure*}[]
\centering
\includegraphics[trim={0 0.15cm 0 0},clip,width=0.8\linewidth]{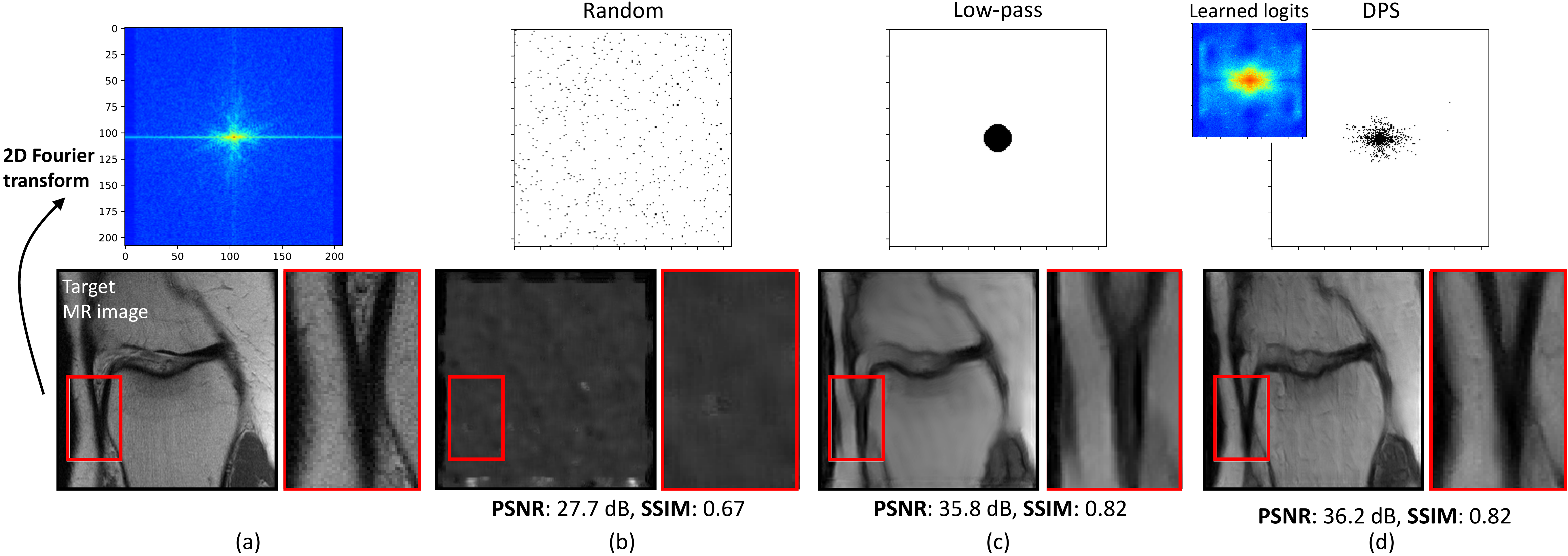}
\caption{\small MR image reconstruction from partial k-space measurements ($N/M=80$) using (b) uniform random sampling, (c) low-pass sampling, and (d) learned sampling by DPS. (d, top inset) learned distribution logits in DPS, expressing beliefs over k-space coefficients. (a) The target image and full k-space, of which only $1.2\%$ of the coefficients are sampled in (b-d).}
\label{fig:MRIrecon}
\end{figure*}
\textbf{Experiment setup}\hspace{0.5cm}
To show the merit of DPS for MRI k-space sampling, we leverage the NYU fastMRI database that contains a vast amount of knee MRI volumes \cite{zbontar2018fastmri}. We selected only the single-coil measurements, and the outer slices of all volumes (which mostly depicted background) were removed. We randomly selected 8000 MRI slices for training, and 2000 and 3000 for validation and testing, respectively. The images were cropped to the central $208\times208$ pixels, and normalized between $0$ and $1$, after which, by means of the 2D fast-Fourier transform, corresponding k-space images $\xvec\in\mathbb{C}^{208 \times 208}$ were generated.   

We compare learned subsampling by DPS to a fixed uniform random or low-pass pattern. For DPS, we here adopt top-$M$ sampling from a single trained distribution to alleviate memory challenges due to the large number of potential k-space coefficients. Results are evaluated through the PSNR and structural similarity index (SSIM) \cite{wang2004image}.

\vspace{0.2cm}

\noindent\textbf{Reconstruction model}\hspace{0.5cm}
Each layer in the MB reconstruction network follows the update rule of \eqref{eqn:ISTAupdate}  (3 unfoldings), with trainable stepsize
$\stepsize^{(k)}$ implemented as a small $3\times 3$ convolutional kernel. The proximal mapping $\prox^{(k)}$ was also learned \cite{mardani2018neural}, by  implementing it as 3 ReLU-activated 2D-convolutional layers with 16 features ($3\times 3$ kernels), followed by a linear 2D-convolutional layer ($3\times 3$ kernels), mapping the output to a single image. \vspace{0.2cm}

\noindent\textbf{Training details}\hspace{0.5cm}
We again adopt the MSE loss, but in addition leverage an adversarial loss to promote visually plausible reconstructions. To that end, we adopt a discriminator network $\Dent$ that is trained to distinguish real from generated MR images \cite{ledig2017photo}, while the sampling and image reconstruction parameters are jointly optimized to maximize discriminator loss. The discriminator consists of 3 2D-convolutional layers with 64 features, $3\times 3$ kernels and a stride of $2$, each followed by a Leaky ReLU activation. The resulting feature maps are then spatially aggregated into a feature vector through global average pooling, followed by dropout ($40\%$). A final fully-connected layer with sigmoid activation then maps the features to a single output probability. 

In addition to the image MSE and adversarial loss, we also penalize the MSE between the discriminator features for generated images and true images. To yield the total training loss, these 3 losses are weighted by multipliers $1$, $5e{-6}$ and $1e{-7}$, respectively.   
%
%
We again use a distinct learning rate for the parameters of the generative sampling model and the other parameters, being 0.01 and $2e{-4}$, respectively. 
The temperature parameter $\tau$ was fixed at $5.0$. Training was performed for 500 epochs with mini-batches containing 8 k-space spectra, using the ADAM solver \cite{kingma2014adam} with settings as in section \ref{sec:TrainingDetailsCaseA}. 
\vspace{0.2cm}

\noindent\textbf{Results}\hspace{0.5cm}
Figure \ref{fig:MRIrecon} shows the obtained MRI reconstructions for an example from the hold-out test set. The k-space was subsampled by a factor 80 (i.e. only $1.2\%$ of the coefficients are sampled), and the results of DPS were compared to non-trained sampling by two baseline methods (random uniform and low-pass). Random uniform sampling (fig. \ref{fig:MRIrecon}b) did not yield plausible reconstructions, lacking sufficient low-frequency coefficients. While low-pass sampling (fig. \ref{fig:MRIrecon}c) resulted in improved reconstructions, it was outperformed by DPS (fig. \ref{fig:MRIrecon}d). The latter achieves an apparent sharpness and detail that is particularly evident from the zoomed areas. Across the full test set, its PSNR value was moreover higher. When inspecting the learned logits we observe that DPS generally favours low frequencies while also distributing samples towards the higher frequencies. 

%
%
%
%
\section{Conclusions}
In this paper we proposed an end-to-end deep learning solution for CS MRI that enables joint optimization of a data-driven task-adaptive subsampling model (deep probabilistic subsampling, DPS) and the parameters of a dedicated model-based signal/image recovery network. We found that learned subsampling yielded improved MR image reconstruction compared to the evaluated non-trained patterns, and showed that the use of a model-based deep reconstruction network facilitates efficient training. Future work includes evaluation of DPS for other MR imaging tasks such as segmentation and classification, as well as its use across a broader range of task-adaptive sampling applications in different fields.  
\label{sec:conclusion}

\bibliographystyle{IEEEbib}
\small{}
\bibliography{strings,refs}

\end{document}